\documentclass[aps,pra,reprint]{revtex4-2}

\usepackage[mathlines]{lineno}
\usepackage{blindtext}
\usepackage[export]{adjustbox}
\usepackage{graphicx,epsfig,epstopdf}
\usepackage{amsmath,amssymb}
\usepackage{esvect}
\usepackage{mathtools}
\usepackage{color}
\usepackage{subfigure}
\usepackage{array}
\usepackage{tabularx}
\usepackage{multirow}
\usepackage{booktabs}
\usepackage{mathtools}
\usepackage{float}
\usepackage{comment}
\newcolumntype{L}[1]{>{\raggedright\arraybackslash}p{#1}}
\newcolumntype{C}[1]{>{\centering\arraybackslash}p{#1}}
\newcolumntype{M}[1]{>{\centering\arraybackslash}m{#1}}
\allowdisplaybreaks

 \usepackage{multirow}
\def\_#1{{\bf #1}}

\def\.{\cdot}

\def\Re{{\rm Re\mit}}

\def\H0{{H_0}}
\def\E0{\eta_0 {H_0}}

\def\=#1{\overline{\overline #1}}

\def\wM{\omega_{\rm M}}
\def\ws{\omega_{\rm s}}

\def\Rload{R_{\rm load}}

\def\Zin{Z_{\rm in}}
\def\mM{m_{\rm M}}

\def\mL{m_{\rm L}}

\def\phiM{\phi_{\rm{M}}}
\def\phiL{\phi_{\rm{L}}}

\def\Vst{V_{\rm s}(t)}
\def\Zs{Z_{\rm s}}

\def\Re{{\rm{Re}}}

\begin{document}

\title{Time-Varying Wireless Power Transfer Systems for Improving Efficiency}
\author{X.~Wang$^{1,2}$}
\author{I.~Krois$^{3}$}
\author{N.~Ha-Van$^{1}$}
\author{M.~S.~Mirmoosa$^{1}$} 
\author{P.~Jayathurathnage$^{1,4}$} 
\author{S.~Hrabar$^{3}$}
\author{S.~A.~Tretyakov$^1$}




\affiliation{$^1$Department of Electronics and Nanoengineering, Aalto University, Finland}
	\affiliation{$^2$Institute of Nanotechnology, Karlsruhe Institute of Technology, Germany}
	\affiliation{$^3$Faculty of Electrical Engineering and Computing, University of Zagreb,  Croatia}
 \affiliation{$^4$Danfoss Drives, Finland}

\begin{abstract}
Conventional wireless power transfer systems are linear and time-invariant, which sets fundamental limitations on their performance, including a tradeoff between transfer efficiency and the level of transferred power. In this paper, we introduce and study a possibility of temporal modulation for inductive wireless power transfer systems and uncover that this tradeoff is avoided as a consequence of varying the 
inductive coupling strength in time. 
Our theoretical analysis reveals that under the optimal modulation depth and phase, the time modulation can
yield a substantial improvement in the WPT efficiency, while
the received power at the load is also improved compared
to the static WPT reference system.
We experimentally demonstrate the concept with a low-frequency system and observe a threefold improvement in efficiency over the reference  static counterpart. 
This technical capability reconciles the inherent tradeoff between the WPT efficiency and transferred power, paving the way for simultaneous advancements in both efficiency and delivered power.
\end{abstract}

\maketitle 


\section{Introduction} 

Wireless power transfer (WPT) is a promising technology that enables transmission of electrical power without the need for wires or cables \cite{1132833,KursScience,Siqi2014, nature2021}. It is desired that both the transfer efficiency and the transferred power are high. However, one of the major challenges of WPT designs is a relatively low efficiency, which can result in significant power losses during transmission. Moreover, there is always a tradeoff between efficiency and transferred power \cite{hui2013critical, Sample2011}. 

Typically, the WPT efficiency is highly influenced by two critical factors: the mutual coupling coefficient $k$ and the quality factor $Q$ of the coils \cite{ChrisMi2016,Li2015Q,Duong2011}. The product  $kQ$ plays a crucial role as it directly characterizes power losses and, consequently, the overall power transfer efficiency of the system~\cite{Prasad2014kQ}. Researchers have made efforts to enhance efficiency by developing various methods to increase these two factors. Recent studies highlighted the use of multiple-coil WPT systems, such as four-coil systems~\cite{KursScience,Sanghoon2011fourcoil},  or relay resonator wireless systems~\cite{hui2013critical}, which leverage the advantages of cascading multiple mutually coupled coils. For example, in~\cite{Sanghoon2011fourcoil}, the four-coil system provides extra mutual coupling  to compensate for the efficiency loss when the transmitting and receiving modules are weakly coupled or placed far apart. Compared to basic two-coil systems, these multiple-coil systems offer substantial improvement in system efficiency. However, the matching impedance condition in these systems still poses an inherent limitation, with the system energy efficiency never exceeding 50\%, which is known as a fundamental limit of WPT systems. In addition, the coupling coefficient, which is directly proportional to the mutual inductance, is mainly dependent on the geometry of the coil structure and the distance between the transmitter and receiver. As a result, there is a limited room for improvement in this critical factor. 
In loosely coupled systems, when the coupling coefficient $k$ is low, it is advisable to use windings with a high $Q$ factor to increase the power transfer efficiency of the WPT systems. In other words, winding technology with low resistance at high-frequency operation is needed for efficient WPT~\cite{RonHui_lookback}. Usually, Litz wire consisting of many thin wire strands is used to reduce the AC winding resistance and improve the $Q$ factor, and therefore enhance the power transfer efficiency~\cite{Deng2016Qfactor, Stein2017Q}.

Nevertheless, conventional WPT systems face inherent limitations in terms of power transfer and efficiency, particularly when the coupling coefficient and quality factor are highly restricted by predetermined working conditions. 
Typically, beyond a certain threshold of efficiency improvement, there is a diminishing return in transferred power, contradicting the very purpose of enhancing efficiency \cite{hui2013critical,zhang2014frequency}. Balancing these two metrics has become the main challenge in designing wireless power transfer systems for specific applications. For example, in some cases, such as in medical implants or electronic devices \cite{agarwal2017wireless}, maximizing efficiency may be more important than delivering a high amount of power. In other cases, like electric vehicle charging, a balance between efficiency and power delivery is sought to ensure practical usability \cite{zakerian2019dynamic}. 
Many studies have aimed to enhance the output power while maintaining the efficiency. This has been achieved through adaptive adjustments of the operating or resonant frequencies \cite{seo2015optimal, mastri2016coupling}. However, dynamically adjusting such a system requires additional information about the operating environment and the efficiency can not be higher than $50\%$.
Essentially, for a fixed WPT system, there is no available method to overcome these  fundamental drawbacks of WPT systems.

Since recently, an alternative way to control the response of electromagnetic systems by varying the effective properties of those systems in time is quickly developing~\cite{AluPEND,TretyakovIAPM}. New possibilities arise because modulation of the system  parameters gives an additional degree of freedom that can be used for enhancing functionalities and breaking conventional limitations~\cite{Monticone23RAPM}. For example, nonreciprocity~\cite{SounasNONR}, frequency translation~\cite{GrbicFTR}, (parametric) amplification~\cite{WangPTC}, wideband impedance matching and increasing bandwidth~\cite{LIALU19,yang2021broadband} indicate only a small part of this enticing and emerging research direction whose roots come back to the previous century (see, e.g., Ref.~\cite{TretyakovIAPM}).

We expect that by incorporating the concept of time modulation as an additional degree of freedom, it is possible to break the fundamental limitations of WPT systems. 
In our previous work \cite{jayathurathnage2021time}, we have predicted that temporally modulating mutual inductance can reconcile the tradeoff between transferred power and efficiency.
In this study, we take a closer look at interactions of time-varying coupling coils, investigating the roles of not only the mutual inductance variations, but all the time-varying inductive components. Based on our theoretical findings, we make  a proof-of-concept experiment to verify the theoretical results. 

The paper is organized as follows: In Section II, we present a developed analytical method that accounts for time-modulated inductance and mutual inductance at the time-varying circuit level. 
In Section III, we investigate the effects of each inductive circuit component of the set of inductively coupled coils, elucidating the physical meanings of time-varying components that contribute to possible enhancements of the WPT performance. 
In Section IV, we report on  a test experimental implementation of a time-modulated WPT system. A simple WPT configuration based on a low-frequency mechanical transducer was implemented as a  proof-of-concept of time-varying mutual inductance enhancing WPT efficiency. 


\section{Theory}


\subsection{Performance indicators of classical wireless power transfer systems}
\begin{figure}[t!]
\centerline{\includegraphics[width= 0.9\columnwidth]{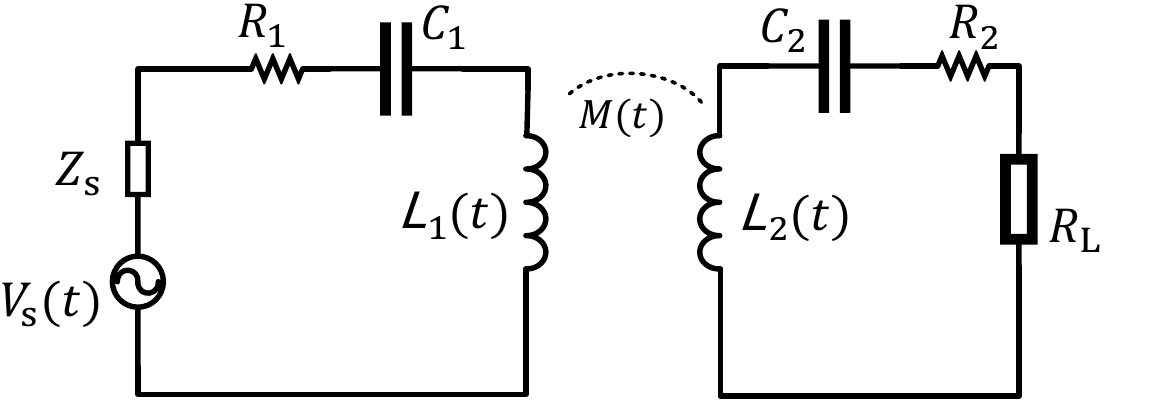}}
\caption{Schematic view of an inductively coupled WPT system where all the inductive components are modulated in  time. }
\label{Fig:WPT_M(t)_circuit}
\end{figure}

To establish a basis for comparison, we initially provide a concise overview of classical WPT systems without any time-modulated components. 
Figure~\ref{Fig:WPT_M(t)_circuit} shows the circuit diagram of a classical inductively coupled WPT system. The system comprises a harmonic voltage  source $\Vst$ with internal impedance $\Zs$, a transmitting resonator (depicted as an $LCR$ circuit $L_1$, $C_1$, and $R_1$), a receiving resonator ($L_2$, $C_2$, and $R_2$), and a resistive load $R_L$. The transmitting resonator is inductively coupled to the receiving resonator, with the level of coupling quantified by the mutual inductance ${M}$. First, we assume that all circuit components remain static, that is, $L_{1,2}(t)$ and $M(t)$ do not depend on time, as in all conventional WPT systems. 

The key performance indicators are the delivered amount of power and the power transfer efficiency, both of which strongly depend on the values of $ {M}$ and $\Rload$. At the resonance frequency $\ws$, the input impedance $\Zin$ seen by the source is given by $\Zin=R_1+(\ws {M})^2/(R_2+ R_L)$ \cite{Bird}. The power transferred  from the source to the load  $P_L$   is maximized  when $\Zin=\Zs^*$ (where $*$ denotes complex conjugate), corresponding to a specific value of $ {M}$ for a given $\Rload$. Deviations from this mutual inductance value result in suboptimal power transfer due to impedance mismatch at the source terminal, leading to reduced delivered power.

The power transfer efficiency $\eta$, primarily influenced by dissipation in the lossy components and the power source. It can be defined as \cite{hui2013critical}
\begin{equation} \label{Eq:WPTpte}
\eta = \frac{P_L}{P_{\rm in}}=\dfrac{P_L }{P_{{R1}}+{P_{ R2}+{P_{s}}+{P_L}}},
\end{equation}
where $P_{R1}$, $P_{R2}$, and $P_{s}$ represent the power losses in the coil resistances $(R_1, R_2)$ and the source resistance $(\Re[\Zs])$. Note that the sum of all the losses and the load power equals to the total input power to the system.

\subsection{Time-modulated inductor and mutual inductors in a circuit}
The fundamental trade-off between maximizing delivered power and  efficiency  \cite{hui2013critical} follows from passivity and time-invariance of the system, meaning that power is supplied only by the primary source and all the circuit components do not change in time. However, if some circuit components are active and/or change in time under action of some external force, this classical limitation can be overcome. Here, we focus on the use of time-varying (parametric) systems. Before we present the results, in this section, we outline the analytical method to analyze time-varying circuits using the  mode-matching method. 

In what follows, we discuss how  time-varying circuit components can be  characterized and how currents and voltages in time-varying circuits can be calculated, with the focus on a time-modulated inductor \cite{jayathurathnage2021time}. 
Let us consider an inductor whose inductance $L$ is periodically modulated in time. We assume that the modulation of inductance is an arbitrary time-periodic function, $L(t)=L(t+T)$, where $T$ is the modulation period. An arbitrary periodic function can be expanded into  Fourier series as 
\begin{equation}
    L(t)=\sum_{m=-\infty}^{+\infty}l_me^{jm\omega_{\rm M}t},\label{eq: L}
\end{equation}
where $\omega_{\rm M} = 2\pi/T$ is the fundamental angular frequency of modulation.
Under a time-harmonic excitation, the time-varying inductance is  a parametric device, where  an infinite number of frequency harmonics is created. The frequency of harmonic $n$ is denoted as $\omega_n=\omega_{\rm s}+n\omega_{\rm M}$ where $\omega_{\rm s}$ is the source frequency.  Therefore, the  current through a time-varying inductor and voltage across it in time domain can be generally written as  sums of all the frequency components:
  \begin{subequations}\label{second:main}
\begin{equation}
    I(t)=\sum_{n=-\infty}^{+\infty}i_ne^{j\omega_nt},\label{eq: I}
\end{equation}
\begin{equation}
    V(t)=\sum_{n=-\infty}^{+\infty}v_ne^{j\omega_nt},\label{eq: V}
\end{equation}
\end{subequations}
where $i_n$ and $v_n$ are the complex amplitudes of the current and voltage harmonics, respectively. We omit the operator Re (real part) in front of the summation sign, for mathematical convenience in the following derivations.
The voltage across an inductor is the time derivative of its magnetic flux:
\begin{equation}
 V(t)=\frac{d\Psi(t)}{dt}=\frac{d[L(t)I(t)]}{dt}, \label{eq: Vt}
\end{equation}
where $\Psi$ is the magnetic flux through the inductor. The above differential equation can be written in integral form after integrating both sides with respect to time: 
\begin{equation}
    \int V(t)dt=L(t)I(t) .\label{eq: int}
\end{equation}
After substituting Eqs.~(\ref{eq: L}), (\ref{eq: I}), and (\ref{eq: V}) into Eq.~(\ref{eq: int}) we have
\begin{equation}
    \sum_{n=-\infty}^{+\infty}\frac{v_n}{j\omega_n}e^{j\omega_nt}= \sum_{n=-\infty}^{+\infty}\sum_{m=-\infty}^{+\infty}i_n l_m e^{j\omega_{n+m}t}. \label{eq: summation1}
\end{equation}
In the right side of Eq.~(\ref{eq: summation1}) we can shift the summation index $n$ by $m$, i.e.,  replace $n$ by $n-m$, so that the left and right sides share the same basis $e^{j\omega_nt}$. After removing the summation over  $n$ at both sides,  the current-voltage relation for harmonic $n$ is found in the form
\begin{equation}
    v_n=\sum_{m=-\infty}^{+\infty}j\omega_n i_{n-m} l_m. \label{eq: vn}
\end{equation}
We can see that due to modulation,  the $n$-th order voltage harmonic is a weighted sum of all the current harmonics $n-m$, where $m\in[-\infty, +\infty]$, with the impedance coefficients $j\omega_nl_m$.
If we consider a finite number of  harmonics from $n=-N$ to $n=+N$ ($N$ is an integer), we can write  $2N+1$ such linear equations as Eq.~(\ref{eq: vn}), and this set of linear equations can be set   in matrix form 
\begin{equation}
    \mathbf{v}=\bar{\bar{Z}}_L \cdot \mathbf{i},
\end{equation}
where
\begin{equation}
\bar{\bar{Z}}_L=\begin{pmatrix}
	j\omega_{-N} l_0 & j\omega_{-N} l_{-1} & \cdots & j\omega_{-N} l_{-2N} \\
j\omega_{{-N+1}} l_1 & j\omega_{{-N+1}} l_0 & \cdots & j\omega_{{-N+1}} l_{1-2N} \\
	\vdots  & \vdots  & \ddots & \vdots  \\
	j\omega_N l_{2N} & j\omega_N l_{2N-1} & \cdots & j\omega_N l_0
	\end{pmatrix}	\label{eq: matrix}
\end{equation}
is the impedance matrix of the time-varying inductance {which should be invertible (the inverse of an impedance matrix is an admittance matrix)}, and 
\begin{subequations}
\begin{align}
     \mathbf{v}=&[v_{-N}, v_{-N+1}, \cdots, v_0, \cdots, v_{N-1}, v_{N}]^{T}\\
     \mathbf{i}=&[v_{-N}, v_{-N+1}, \cdots, v_0, \cdots, v_{N-1}, v_{N}]^{T}   
\end{align}
\end{subequations}
denote the voltage and current arrays, respectively. We can see that in time-modulated circuits the currents and voltages are characterized by \textit{arrays} containing complex amplitudes of all the considered frequency components, and the impedances are \textit{matrices}. 
Similarly, a time-varying mutual inductance can be represented as such matrix, which we denote as  $\bar{\bar{Z}}_{M}$

In such multi-frequency systems, the impedances of static components should also be characterized by matrices. In contrast to time-varying inductances, static components do not create mutual coupling between different harmonics of voltages and currents. Therefore, their impedance matrix representations are purely diagonal. The diagonal terms are the impedance values at the frequencies  of the corresponding harmonic components. We denote the impedance matrices of components $Z_s$, $C_1$, $R_1$, $C_2$, $R_2$,  and $R_L$ as $\bar{\bar{Z}}_s$, $\bar{\bar{Z}}_{C1}$, $\bar{\bar{Z}}_{R1}$, $\bar{\bar{Z}}_{C2}$, $\bar{\bar{Z}}_{R2}$, and $\bar{\bar{Z}}_{RL}$.

\begin{figure}[h!]
\centerline{\includegraphics[width= 0.8\columnwidth]{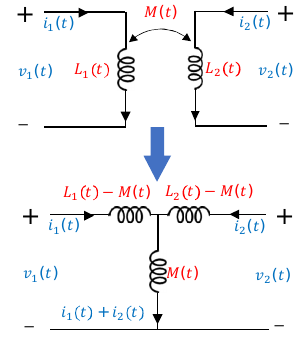}}
\caption{Equivalence of time-varying magnetically coupled coils to time-varying T-circuits.}
\label{Fig:t-circuit}
\end{figure}

Next, we calculate the current and voltage of the circuit considering all the harmonics. Before doing this, we transform the inductive coupling circuit into an equivalent T-circuit, as shown in Fig.~\ref{Fig:t-circuit}. Such transformation is usually performed for static systems. In Appendix, we prove that this equivalence is also valid if the coupling transformer is varying in time ($L_{1,2}(t)$ and $M(t)$). In this way, 
the calculation of currents and voltages through or across each component can follow the traditional way, simply replacing the scalar parameter values with the corresponding matrices. First, we calculate the current in the transmitter coil, $\mathbf{i}_s$. According to the equivalent T-circuit in Fig.~\ref{Fig:t-circuit}, the total impedance seen from the source can be calculated as
\begin{align*}
\bar{\bar{Z}}_{\rm tot}&=\bar{\bar{Z}}_s+\bar{\bar{Z}}_{R1}+\bar{\bar{Z}}_{C1}+\bar{\bar{Z}}_{L1}-\bar{\bar{Z}}_{M}\\
&+\left[\left(\bar{\bar{Z}}_{R2}+\bar{\bar{Z}}_{C2}+\bar{\bar{Z}}_{RL}+\bar{\bar{Z}}_{L2}-\bar{\bar{Z}}_{M}\right)^{-1}+\bar{\bar{Z}}_{M}^{-1}\right]^{-1}.
\end{align*}
The current can be calculated as the dot multiplication of the  admittance matrix (inverse of the impedance matrix) and the voltage vector:
\begin{equation}
    \mathbf{i}_s=\bar{\bar{Z}}_{\rm tot}^{-1}\cdot \mathbf{v}_s.
\end{equation}
Similarly,  we can calculate the voltage and current in the load impedance, $\mathbf{i}_L$ and $\mathbf{v}_L$:
\begin{equation}
    \mathbf{i}_L=(\bar{\bar{Z}}_{L2}-\bar{\bar{Z}}_{M}+\bar{\bar{Z}}_{R2}+\bar{\bar{Z}}_{C2}+\bar{\bar{Z}}_{RL})^{-1}\cdot \mathbf{v}_M,
    \end{equation}
where $\mathbf{v}_M=\mathbf{v}_s-(\bar{\bar{Z}}_s+\bar{\bar{Z}}_{C1}+\bar{\bar{Z}}_{R1}+\bar{\bar{Z}}_{L1}-\bar{\bar{Z}}_{M})\cdot \mathbf{i}_s $. Then, the voltage across the load can be found as $\mathbf{v}_L=\bar{\bar{Z}}_{RL}^{-1}\cdot \mathbf{i}_L$.
Knowing all the currents and voltages in the circuit, we can calculate the efficiency by Eq.~(\ref{Eq:WPTpte})

\section{Efficiency analysis of time-modulated WPT systems}
\begin{figure}[h]
	\centerline{\includegraphics[width= 0.9\columnwidth]{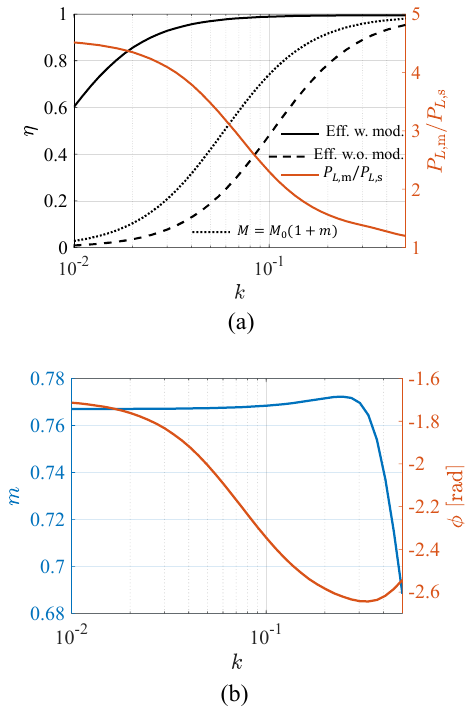}}
 		\caption{(a) Power transfer efficiency and the output power at the load with and without modulation,  in terms of the coupling coefficient. The modulation parameters for each value of $k$ are optimized independently.  (b) The optimized modulation parameters for each value of $k$. }
 		\vspace{-10pt}
 		\label{Fig:eta-k}
\end{figure}

In this section, we analyse the effects of time-modulation of inductances on the performance of  WPT systems when the modulation is made at twice the  frequency of the power source. Here, we consider modulation of the self-inductances of the Tx and Rx coils (i.e., $L_1(t)$ and $L_2(t)$) and of the mutual inductance (i.e., $M(t)$) as expressed in the following relations:
\begin{eqnarray}
	L_{\{1,2\}}(t)&=&L_{0}\Big[1 + \mL{_{\{1,2\}}} \cos(\wM t+\phiL{_{\{1,2\}}})\Big]\\
	M(t)&=&M_0\Big[1 + \mM \cos(\wM t+\phiM)\Big].
\end{eqnarray}
Here, $m_{L\{1, 2\}}$ and $m_{\rm M}$ are the modulation depths of the self-inductance and mutual inductance, respectively, and $\phi_{L\{1, 2\}}$ and $\phi_{\rm M}$ are the modulation phases of self and mutual inductances, respectively. These parameters in principle can be controlled independently. Here, for simplicity of the analysis, we consider the self and mutual inductances to be modulated with the same modulation depth and phase, i.e.,  $m_{L\{1, 2\}}=m_{\rm M}=m$ and $\phi_{L\{1, 2\}}=\phi_{\rm M}=\phi$. For this modulation case, we study possibilities of using such time-varying circuits for  improvement of the device performance. As an example, we assume  $Z_{\rm s}=2~\Omega$, $R_L=20~\Omega$, $R_1=R_2=0.1~\Omega$, $C_1=C_2=25.33$~nF, $L_0=100~\mu$H, $\omega_{\rm s}=2\pi\times100$~kHz, $\omega_{\rm M}=2\omega_{\rm s}$. 
Figure~\ref{Fig:eta-k}
shows the efficiency and the power received by the load with respect to the coupling coefficient $k$, for two cases, with and without modulation. It can be seen that with modulation the wireless transfer efficiency can be improved significantly. For small coupling coefficients ($k\approx0.01$), the efficiency is near zero when the coil is not modulated. Strikingly, with modulation, the efficiency improves significantly ($\eta>0.6$). It is also clear that the efficiency for the modulated case is always larger than for the corresponding static system in a large range of $k$. The increase in efficiency is accompanied with an increase of the power transferred to the load. One can see that the load power is always larger than that in the static system, i.e., $P_{L,m}/P_{L,s}>1$ for all values of $k$ where $P_{L,m}$ and $P_{L,s}$ represent the received power at the load for modulated and static cases, respectively. The results in Fig.~\ref{Fig:eta-k}(a) are obtained via optimization of $m$ and $\phi$. The optimized modulation parameters for each value of $k$ are displayed in Fig.~\ref{Fig:eta-k}(b). It should be noted that the improvement of efficiency under time modulation is not caused by the enhanced mutual inductance within one half of the modulation period, $M_0(1+m)$. This can be simply seen in Fig.~\ref{Fig:eta-k}(a), where the efficiency of a static WPT system with $M=M_0(1+m)$ is plotted (see black dotted line).

  \begin{figure*}
	\centerline{\includegraphics[width= 2.0\columnwidth]{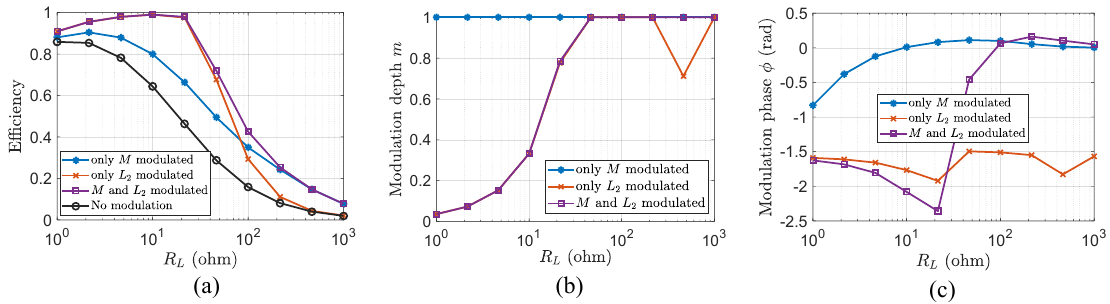}}
 		\caption{(a) Optimized efficiencies, (b) modulation depths, and (c) modulation phases for different values of the load resistance.  }
 		\vspace{-10pt}
 		\label{Fig:comparison}
\end{figure*}

Next, we investigate the impact of individual component modulation on the efficiency of the WPT system. We begin by examining the effect of modulating $L_1$, which induces an effective negative resistance in the primary branch at the optimal phase angle \cite{jayathurathnage2021time}. Basically, in this case the time-modulated inductance acts as a parametric amplifier in the transmitter circuit. This results in an increase of the current in the transmitter and receiver loops, thereby boosting the output power. However, although the modulated $L_1$ configuration yields higher power, it does not necessarily enhance efficiency. In fact, the modulation of  only $L_1$ increases losses in proportion to the power increase, thus, no efficiency improvement is achieved. On the other hand, modulating $L_2$ emulates a negative resistance in the receiver loop, resulting in an increase in the power delivered to the receiver without incurring additional losses in the transmitter coil or in the source resistance. This leads to a significant improvement of efficiency. Similarly, the modulation of $M$ pumps energy directly to the receiver loop at the optimal phase angle, which also contributes to the efficiency enhancement of the WPT system. Obviously, there are inevitable additional losses in the modulation circuits, but since these circuits do not carry significant power (they only modulate the reactive components), these additional losses are expected to be not significant in comparison to dissipation in the coils and in the main power source.

In order to evaluate the effects of time modulations of separate inductors, we fix the coupling coefficient to $k=0.1$ and examine the efficiency improvement as a function of the load resistance (presented in Fig.~\ref{Fig:comparison}). Here, we consider three cases, modulating only $M$, $L_2$, and $M$ and $L_2$ simultaneously.  Figure~\ref{Fig:comparison}(a)  shows the efficiency in these scenarios. Note that the modulation phase is optimized to achieve maximum efficiency at each load. We see that the mutual inductance modulation (modulating only $M$) has a positive effect on the efficiency improvement, for all values of the considered load resistance ($1~\Omega<R_L<1000~\Omega$).  Modulation of only $L_2$ has a much more positive effect. Therefore, modulating $L_2$ and $M$ simultaneously (with the same modulation phase) is potentially the optimal way for efficiency improvement.  The modulation phases and amplitudes for the three modulation scenarios are displayed in Fig.~\ref{Fig:comparison}(b,c) that are obtained from optimization.  For all three cases, we confirm that the received power at the load is always larger than in the static case.

\section{Experimental study}

\subsection{Experiment setup}

\begin{figure}[h]
	\centerline{\includegraphics[width= 1\columnwidth]{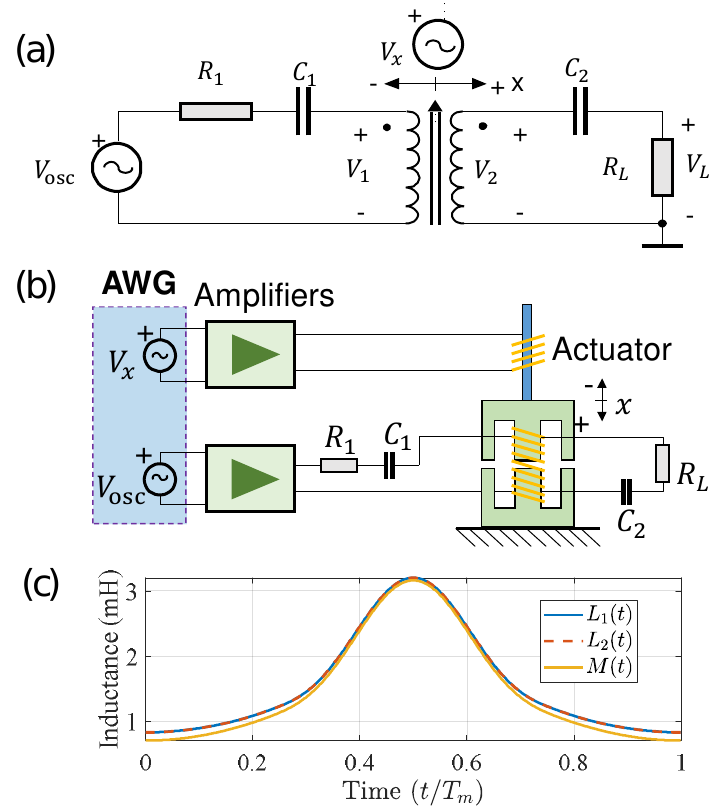}}
 		\caption{Experimental WPT demonstrator based on a mechanically time-varying mutual inductance. (a) Schematic diagram. (b) Simplified construction. (c) Modulation function  of $L_1$, $L_2$, and $M$ in one temporal period. $T_{\rm M}=2\pi/\omega_{\rm M}$ is the modulation period.   }
 		\vspace{-10pt}
 		\label{Fig:exp}
\end{figure}

It is important to note that practical implementation of any electrically-tunable inductance is very challenging. There are some attempts for creation of tunable inductors based on MEMS \cite{el2010liquid}, magnetoelectric materials \cite{lou2009electrostatically}, and non-Foster-based systems \cite{vorobev2017active}. However, all these systems are still in the very early phase of the research. To the best of our knowledge, there are no works on tuning mutual inductance. Therefore, for the experimental verification of our concept we used a simple mechanical system to modulate the mutual inductance between two coils.
Specifically,  a wireless power transfer demonstrator based on a  low-frequency mechanical realization of a time-varying mutual inductance has been designed (Fig.~\ref{Fig:exp}). We focus our efforts on realization of time-modulated mutual inductance, because modulation of ordinary inductance creates better understood parametric amplification, while it appears that the effects of modulations of coupling have not been studied before.  

A schematic diagram of the system (see Fig.~\ref{Fig:exp}(a)) resembles a classical WPT system that transfers energy from the source ($V_{\rm osc}$) to a load ($R_L$) via a resonant pair of inductively coupled coils, forming a transformer. However, instead of having a  fixed mutual inductance (and fixed transformation ratio), this transformer is a time-dependent device.  It is based on a time-varying mutual inductance controlled by a modulation signal $V_x$ (a pump), as is illustrated in Fig.~\ref{Fig:exp}(b).  
 
 \begin{figure}[h]
	\centerline{\includegraphics[width= 1\columnwidth]{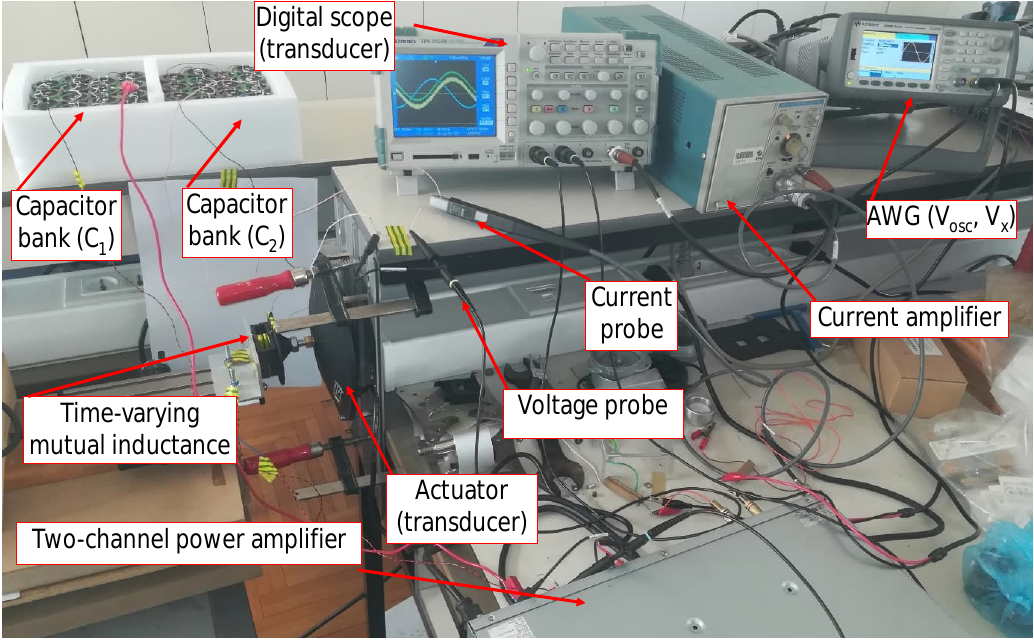}}
 		\caption{Photo of the experimental setup.}
 		\vspace{-10pt}
 		\label{Fig:photo}
\end{figure}

The transformer comprises two commercial E52 ferrite cores with a mutual air gap. The primary and secondary coils have 62 turns of 0.6~mm diameter copper wire, wound in a bifilar fashion. The coils were wound around both central cores, one on top of the other. One side of the ferrite core is fixed to a solid base, while the second one is mounted onto an electromechanical transducer. The mechanical construction of the ‘moving joint’ was very challenging since even very small mechanical vibrations (in the order of a tenth of a millimetre) can affect the length of the air gap and, therefore, the mutual inductance. For this reason,  a precise  two-axes mechanical positioner was used for accurate adjustment of desired core locations. The transducer is driven by a pump signal (the first output of a two-channel arbitrary waveform generator, AWG) amplified with the help of an in-house  two-channel class~D power amplifier. Thus, mechanical modulation of the mutual inductance is achieved by movement of one of the  ferrite cores, i.e., by varying the length of the air gap.  The second output of the arbitrary waveform generator was connected to the second channel of the class~D power amplifier, feeding the primary coil of a transformer while the secondary coil was loaded by a resistive load. Thanks to  high-efficient  D-class operation, no significant thermal problems were noted even with very high output current levels (up to 3~A), needed for excitation of the mechanical transducer. 

The circuit parameters in the experiment are as following: $R_1=R_2=0.3~\Omega$, $C_1=C_2=88$~mF, $R_L=4.7~\Omega$, $V_{\rm osc}=5$~V with the source resistance $Z_{\rm s}=5.6~\Omega$. The  modulation (pump) frequency is limited by the mechanical self-resonance of the system, which is about 110~Hz. In
order to avoid parasitic effects caused by the mechanical self-resonance,   the modulation frequency should be lower than the self-resonance frequency. Therefore, we chose the modulation frequency of $\omega_{\rm M}=2\pi \times 90$~Hz and the source frequency of $\omega_{\rm s}=2\pi \times 45$~Hz. The modulation profiles for $L_1(t)$, $L_2(t)$, and $M(t)$ are shown in Fig.~\ref{Fig:exp}(c). Furthermore, we  paid special attention to achieving the required phase shift between the input and pump signals. The problem is that the phase shift set on the AWG does not match the actual phase shift between the two signals, due to different delays in the two channels of the power amplifier and the different lengths of the connecting cables. Therefore, we set a zero phase at the AWG and checked the phase difference between the signals at the transducer and at the primary winding of a transformer with an  oscilloscope. Then we introduced an additional phase shift required to compensate for the delay difference. This correction is very effective for the referent phase shift (zero degree), while its accuracy decreases for other phase shifts. Another practical problem is the unknown output impedance of the power amplifier (in the theoretical part this impedance was assumed to be zero). We do not know the exact value of this impedance, but a ``rule of thumb'' estimate gives a value of about 0.1~$\Omega$ to 0.3~$\Omega$. Therefore, we calculated the associated voltage attenuation ratios and compensated for them by making appropriate adjustments to the gains of the oscilloscope channels. After all these initial calibration procedures were completed, we began the measurements that will be explained in the next section.


\begin{figure*} \centerline{\includegraphics[width= 2.0\columnwidth]{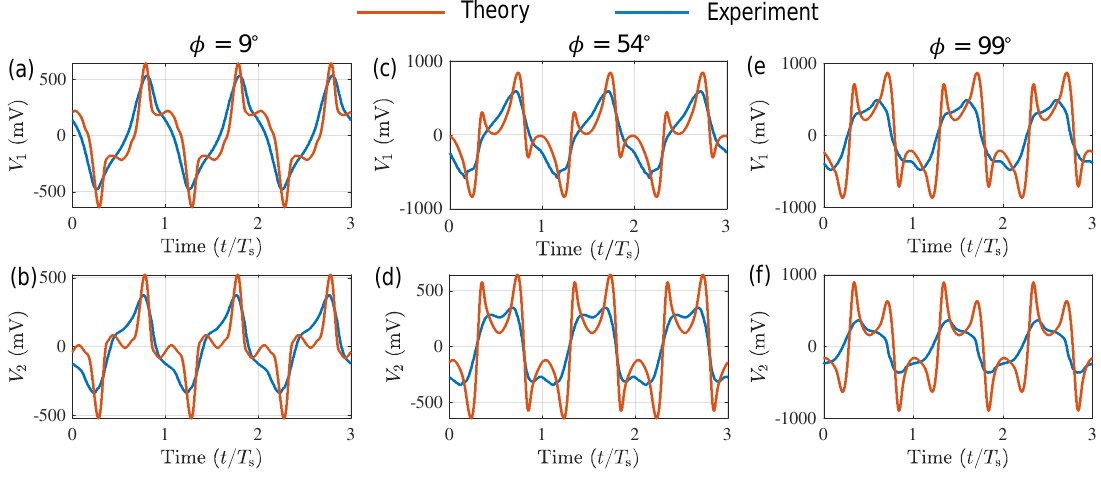}} 
\caption{(a)(b)(c) The theoretical (orange line) and measured (blue line) voltages at the primary coil $V_1$ and (d)(e)(f) the secondary coil $V_2$ for the modulation phases of $\phi=9^\circ$, $\phi=54^\circ$, and $\phi=99^\circ$.} 
\vspace{-10pt} 
\label{Fig:measurement} 
\end{figure*}


\subsection{Results}
The theoretical and measured waveforms of $V_1$ and $V_2$ are compared in Fig.~\ref{Fig:measurement}. We consider three scenarios with the modulation phases $\phi=9^\circ$, $\phi=54^\circ$, and $\phi=99^\circ$. It can be seen that in all three scenarios the voltage profile in the time domain is not sinusoidal, which is a clear evidence that higher-order harmonics are generated. The measured signals are quite similar to those obtained from the theory. There are slight differences that are more pronounced at larger phase delays. This is a direct consequence of the limitations of the experimental system described in the previous section. Nevertheless, the obtained results clearly support the conclusions from the developed theory. This is also evident from the table showing the measured power received by the load with and without modulation. It can be seen that the efficiency is significantly higher for some modulation phases than for the static system, for example, $\phi=54^\circ$ and $\phi=99^\circ$ in our experiment. Of course, like in any parametric system, this time-varying system is phase dependent. One must synchronize the phases of  the signal and modulation in order to obtain an improved efficiency. 


\begin{table}[]
\begin{tabular}{|llc|}
\hline
\multicolumn{3}{|c|}{WPT Efficiency}                                                                                        \\ \hline
\multicolumn{1}{|l|}{Modulation phase} & \multicolumn{1}{l|}{with modulation} & \multicolumn{1}{l|}{without modulation} \\ \hline
\multicolumn{1}{|l|}{$\phi=9^\circ$}                & \multicolumn{1}{l|}{8.13\%}          & \multirow{3}{*}{9.28\%}                 \\ \cline{1-2}
\multicolumn{1}{|l|}{$\phi=54^\circ$}               & \multicolumn{1}{l|}{19.8\%}          &                                         \\ \cline{1-2}
\multicolumn{1}{|l|}{$\phi=99^\circ$}               & \multicolumn{1}{l|}{27.73\%}         &                                         \\ \hline
\end{tabular} \caption{Measured efficiency with and without modulation for three different modulation phases.}
\end{table}

\section{Conclusion}
In summary, this paper introduces an inductive wireless power transfer system with coupling inductance varying periodically over time. The study finds that by varying the inductance of the coupled coils and the mutual inductance, the WPT system's efficiency can be increased significantly. At the same time, the power received by the load is also improved as compared to the reference static WPT system. This technique overcomes the fundamental tradeoff between efficiency and delivered power, improving the transferred power and transfer efficiency simultaneously. 
To validate this conclusion, a conceptual experiment was conducted, demonstrating that modulation of the coil inductance and mutual inductance can lead to a significant improvement in power transfer efficiency. The experimental results confirm the theoretical conclusions drawn from the study.


\bibliographystyle{Bibliography/IEEEtranTIE}
\bibliography{Bibliography/IEEEabrv,Bibliography/BIB_xx-TIE-xxxx}\ 


  \appendix*
\section{Validity of the T-circuit interpretation for time-varying coupled coils}
It is known that two inductively coupled coils can be considered equivalent to a T-circuit with inductances $L_1-M$, $L_2-M$, and $M$. This equivalence is usually derived for static components, meaning that $L_1$, $L_2$, and $M$ are time-invariant. Next, we demonstrate that  the T-circuit interpretation of mutual inductance is also valid when the components are varying as arbitrary functions of time. 

In the time domain, the voltage at the two sides of a coil (see Fig.~\ref{Fig:t-circuit}(top)) can be expressed as
\begin{equation}
    v_1(t)=\frac{ d[L_1(t)i_1(t)]}{dt}+\frac{ d[M(t)i_2(t)]}{dt},
\end{equation}
\begin{equation}
    v_2(t)=\frac{ d[L_2(t)i_2(t)]}{dt}+\frac{ d[M(t)i_1(t)]}{dt}.
\end{equation}
These two equations can be rearranged as
\begin{equation}
    v_1(t)=\frac{ d\left\{[L_1(t)-M(t)]i_1(t)\right\}}{dt}+\frac{ d\left\{M(t)[i_1(t)+i_2(t)]\right\}}{dt},
\end{equation}
\begin{equation}
    v_2(t)=\frac{ d\left\{[L_2(t)-M(t)]i_2(t)\right\}}{dt}+\frac{ d\left\{M(t)[i_1(t)+i_2(t)]\right\}}{dt}.
\end{equation}
One can see that the above two equations are exactly the voltage-current relation of a T-circuit (see Fig.~\ref{Fig:t-circuit}(bottom)). Therefore, time-varying magnetically coupled coils can still be considered equivalent to a T-circuit. This property brings convenience for the analysis of time-varying coils.

\end{document}